\documentclass[12pt,english]{article}
\usepackage[T1]{fontenc}
\usepackage[latin9]{inputenc}
\usepackage{graphicx}
\usepackage{amssymb}

\makeatletter

\providecommand{\tabularnewline}{\\}

\usepackage{a4wide}
\usepackage{psfrag}
\makeatletter

\@addtoreset{equation}{section}

\@addtoreset{figure}{section}

\@addtoreset{table}{section}
\makeatother 

\makeatother

\usepackage{babel}

\begin{document}

\title{\begin{flushright}{\normalsize ITP-Budapest Report No. 652}\end{flushright}\vspace{1cm}Breather
boundary form factors in sine-Gordon theory}

\author{M. Lencsés\\
 Budapest University of Technology and Economics\\
 and\\
 G. Takács\\
 HAS Theoretical Physics Research Group\\
 1117 Budapest, Pázmány Péter sétány 1/A, Hungary}

\date{9th June 2011}
\maketitle
\begin{abstract}
A previously conjectured set of exact form factors of boundary exponential
operators in the sinh-Gordon model is tested against numerical results
from boundary truncated conformal space approach in boundary sine-Gordon
theory, related by analytic continuation to sinh-Gordon model. We
find that the numerical data strongly support the validity of the
form factors themselves; however, we also report a discrepancy in
the case of diagonal matrix elements, which remains unresolved for
the time being.
\end{abstract}

\section{Introduction}

The investigation of integrable boundary quantum field theories started
with the seminal work of Ghoshal and Zamolodchikov \cite{GZ}, who
set up the boundary R-matrix bootstrap, which makes possible the determination
of the reflection matrices and provides complete description of the
theory on the mass shell.

For the calculation of correlation functions, matrix elements of local
operators between asymptotic states have to be computed. In a boundary
quantum field theory there are two types of operators, the bulk and
the boundary operators, where their names indicate their localization
point. The boundary form factor program for calculating the matrix
elements of local boundary operators between asymptotic states was
initiated in \cite{bffprogram}. The validity of form factor solutions
was checked in the case of the boundary scaling Lee-Yang model by
calculating the two-point function using a spectral sum and comparing
it to the prediction of conformal perturbation theory. In \cite{bffcount}
the spectrum of independent form factor solutions in the scaling Lee-Yang
model and the sinh-Gordon model was compared to the boundary operator
content of the ultraviolet boundary conformal field theory and a complete
agreement was found. It is also possible to compare form factors to
matrix elements of local operators evaluated directly from the boundary
quantum field theory in a non-perturbative framework. For periodic
boundary conditions, this was developed in \cite{fftcsa1,fftcsa2};
the extension to boundary form factors was obtained in \cite{bfftcsa}
and used to verify the results of the form factor bootstrap in the
scaling Lee-Yang model by comparison to boundary truncated conformal
space approach. 

Further solutions of the boundary form factor axioms were constructed
and their structure was analyzed for the sinh-Gordon theory at the
self-dual point in \cite{ca1}, and for the $A_{2}$ affine Toda field
theory in \cite{ca2}. One of the present authors constructed a solution
for boundary exponential operators in sinh-Gordon theory \cite{bshgff},
and the solution was checked by computing the conformal dimensions
and vacuum expectation values of the fields in a cumulant expansion
ordered by powers of the coupling constant, which can be compared
to known exact results. However, the conformal dimension essentially
tests only the part constructed out of the bulk form factors, and
the expectation value was checked only to the lowest nontrivial order
in the coupling constant.

Our aim in this paper is to provide a detailed non-perturbative verification
of the solution presented in \cite{bshgff}: using the ideas of \cite{bfftcsa}
we aim to compare the form factors to numerically computed finite
volume matrix elements. However, this cannot be performed directly
in the sinh-Gordon theory as we have no way to construct a truncated
conformal space in this case. Fortunately, it is easy to argue that
(at least to all order of perturbation theory) an analytic continuation
to sine-Gordon model should work, and for this model working truncated
conformal space program was developed in \cite{neumann,finitesizebsg}.
This forms the basis of the present work.

The paper is structured as follows. In Section 2 the boundary form
factors of exponential operators in the sinh-Gordon model are recalled
(in doing so some unfortunate typos in the paper \cite{bshgff} are
also fixed). In Section 3 these form factors are analytically continued
to obtain breather form factors in sine-Gordon theory, and summarize
briefly the necessary ingredients to obtain predictions finite volume
matrix elements. Section 4 contains our numerical analysis, and we
present our conclusions in section 5.

\section{Boundary form factors in the sinh-Gordon model}

\subsection{Boundary sinh-Gordon model}

The sinh-Gordon theory in the bulk is defined by the Lagrangian density
\[
\mathcal{L}=\frac{1}{2}(\partial_{\mu}\Phi)^{2}-\frac{m^{2}}{b^{2}}(\cosh b\Phi-1)\]
It can be considered as the analytic continuation of the sine-Gordon
model for imaginary coupling. The S-matrix of the model is\begin{equation}
S(\theta)=-\left(1+\frac{B}{2}\right)_{\theta}\left(-\frac{B}{2}\right)_{\theta}=\left[-\frac{B}{2}\right]_{\theta}\qquad;\quad B=\frac{2b^{2}}{8\pi+b^{2}}\label{eq:sinh_Smatrix}\end{equation}
where \[
(x)_{\theta}=\frac{\sinh\frac{1}{2}\left(\theta+i\pi x\right)}{\sinh\frac{1}{2}\left(\theta-i\pi x\right)}\quad,\quad[x]_{\theta}=-(x)_{\theta}(1-x)_{\theta}=\frac{\sinh\theta+i\sin\pi x}{\sinh\theta-i\sin\pi x}\]
 The minimal bulk two-particle form factor belonging to this S-matrix
is \cite{FMS}\begin{equation}
f(\theta)=\mathcal{N}\exp\left[8\int_{0}^{\infty}\frac{dx}{x}\sin^{2}\left(\frac{x(i\pi-\theta)}{2\pi}\right)\frac{\sinh\frac{xB}{4}\sinh(1-\frac{B}{2})\frac{x}{2}\sinh\frac{x}{2}}{\sinh^{2}x}\right]\label{eq:fmin}\end{equation}
where \begin{equation}
\mathcal{N}=\exp\left[-4\int_{0}^{\infty}\frac{dx}{x}\frac{\sinh\frac{xB}{4}\sinh(1-\frac{B}{2})\frac{x}{2}\sinh\frac{x}{2}}{\sinh^{2}x}\right]\label{eq:NN}\end{equation}
It satisfies $f(\theta,B)\rightarrow1$ as $\theta\rightarrow\infty$,
and approaches its asymptotic value exponentially fast.

Sinh-Gordon theory can be restricted to the negative half-line with
the following action\begin{eqnarray}
\mathcal{A} & = & \int_{-\infty}^{\infty}dt\int_{-\infty}^{0}dx\left[\frac{1}{2}(\partial_{\mu}\Phi)^{2}-\frac{m^{2}}{b^{2}}(\cosh b\Phi-1)\right]\label{eq:boundaryaction}\\
 &  & +\int_{-\infty}^{\infty}dtM_{0}\left[\cosh\left(\frac{b}{2}(\Phi(0,t)-\Phi_{0})\right)-1\right]\nonumber \end{eqnarray}
which maintains integrability \cite{GZ}. The corresponding reflection
factor depends on two continuous parameters and can be written as
\cite{shGEF}\begin{equation}
R(\theta)=\left(\frac{1}{2}\right)_{\theta}\left(\frac{1}{2}+\frac{B}{4}\right)_{\theta}\left(1-\frac{B}{4}\right)_{\theta}\left[\frac{E-1}{2}\right]_{\theta}\left[\frac{F-1}{2}\right]_{\theta}\label{eq:shgrefl}\end{equation}
It can be obtained as the analytic continuation of the first breather
reflection factor in boundary sine-Gordon model which was calculated
by Ghoshal in \cite{ghoshal}. The relation of the bootstrap parameters
$E$ and $F$ to the parameters of the Lagrangian is known both from
a semi-classical calculation \cite{shGEF,corrigan} and also in an
exact form in the perturbed boundary conformal field theory framework
\cite{sinG_uvir}.

\subsection{Boundary form factors in sinh-Gordon theory \label{sub:Boundary-form-factors}}

Here we recall the results of \cite{bshgff}, but with a few typos
corrected in the expression of the form factor polynomials in (\ref{eq:Bequ})
and (\ref{eq:Apolynomials}). For a local operator $\mathcal{O}(t)$
localized at the boundary (located at $x=0$, and parametrized by
the time coordinate $t$) the form factors are defined as\begin{eqnarray*}
\,_{out}\langle\theta_{1}^{'},\theta_{2}^{'},\dots,\theta_{m}^{'}\vert\mathcal{O}(t)\vert\theta_{1},\theta_{2},\dots,\theta_{n}\rangle_{in} & =\\
 &  & \hspace{-2cm}F_{mn}^{\mathcal{O}}(\theta_{1}^{'},\theta_{2}^{'},\dots,\theta_{m}^{'};\theta_{1},\theta_{2},\dots,\theta_{n})e^{-imt(\sum\cosh\theta_{i}-\sum\cosh\theta_{j}^{'})}\end{eqnarray*}
for $\theta_{1}>\theta_{2}>\dots>\theta_{n}>0$ and $\theta_{1}^{'}<\theta_{2}^{'}<\dots<\theta_{m}^{'}<0$,
using the asymptotic $in/out$ state formalism introduced in \cite{BBT}.
They can be extended analytically to other values of rapidities. With
the help of the crossing relations derived in \cite{bffprogram} all
form factors can be expressed in terms of the elementary form factors\[
\,_{out}\langle0\vert\mathcal{O}(0)\vert\theta_{1},\theta_{2},\dots,\theta_{n}\rangle_{in}=F_{n}^{\mathcal{O}}(\theta_{1},\theta_{2},\dots,\theta_{n})\]
The general form factor solution can be written as \cite{bffprogram}\begin{equation}
F_{n}(\theta_{1},\theta_{2},\dots,\theta_{n})=H_{n}\frac{Q_{n}(y_{1},y_{2}\dots,y_{n})}{\prod_{i}y_{i}\,\prod\limits _{i<j}(y_{i}+y_{j})}\prod_{i=1}^{n}r(\theta_{i})\prod_{i<j}f(\theta_{i}-\theta_{j})f(\theta_{i}+\theta_{j})\label{eq:GenAnsatz}\end{equation}
where\begin{equation}
y=2\cosh\theta\label{eq:ydef}\end{equation}
The $Q_{n}$ are symmetric polynomials of its variables, and the minimal
one-particle boundary form factor is given by\begin{equation}
r(\theta)=\frac{i\sinh\theta}{(\sinh\theta-i\sin\gamma)(\sinh\theta-i\sin\gamma')}u(\theta,B)\quad,\quad\gamma=\frac{\pi}{2}(E-1)\quad\gamma'=\frac{\pi}{2}(F-1)\label{eq:rmin}\end{equation}
where \begin{eqnarray}
u(\theta) & = & \exp\Bigg\{\int_{0}^{\infty}\frac{dt}{t}\left[\frac{1}{\sinh\frac{t}{2}}-2\cosh\frac{t}{2}\cos\left[\left(\frac{i\pi}{2}-\theta\right)\frac{t}{\pi}\right]\right]\times\nonumber \\
 &  & \frac{\sinh\frac{tB}{4}+\sinh\left(1-\frac{B}{2}\right)\frac{t}{2}+\sinh\frac{t}{2}}{\sinh^{2}t}\Bigg\}\label{eq:ufunc}\end{eqnarray}
and \begin{equation}
H_{n}=\left(\frac{4\sin\pi B/2}{f(i\pi)}\right)^{n/2}\label{eq:Hn}\end{equation}
is a convenient normalization factor. The polynomials $Q_{n}$ satisfy
the following recursion relations: \begin{eqnarray}
\mathcal{K}: &  & Q_{2}(-y,y)=0\nonumber \\
 &  & Q_{n+2}(-y,y,y_{1},\dots,y_{n})=\nonumber \\
 &  & (y^{2}-4\cos^{2}\gamma)(y^{2}-4\cos^{2}\gamma')P_{n}(y|y_{1},\dots,y_{n})\, Q_{n}(y_{1},\dots,y_{n})\quad\mbox{for }n>0\nonumber \\
\mathcal{B}: &  & Q_{1}(0)=0\nonumber \\
 &  & Q_{n+1}(0,y_{1},\dots,y_{n})=\nonumber \\
 &  & 4\cos\gamma\cos\gamma'B_{n}(y_{1},\dots,y_{n})\, Q_{n}(y_{1},\dots,y_{n})\quad\mbox{for }n>0\label{eq:sinhgrecursions}\end{eqnarray}
where\begin{eqnarray}
B_{n}(y_{1},\dots,y_{n}) & = & \frac{1}{4\sin\frac{\pi B}{2}}\left(\prod_{i=1}^{n}\left(y_{i}-2\sin\frac{\pi B}{2}\right)-\prod_{i=1}^{n}\left(y_{i}+2\sin\frac{\pi B}{2}\right)\right)\nonumber \\
 & = & -\sum_{l=0}^{\left[\frac{n-1}{2}\right]}\left(2\sin\frac{\pi B}{2}\right)^{2l}\sigma_{n-1-2l}\label{eq:Bequ}\end{eqnarray}
and \begin{equation}
P_{n}(y|y_{1},\dots y_{n})=\frac{1}{2(y_{+}-y_{-})}\left[\prod_{i=1}^{n}(y_{i}-y_{-})(y_{i}+y_{+})-\prod_{i=1}^{n}(y_{i}+y_{-})(y_{i}-y_{+})\right]\label{eq:Pequ}\end{equation}
with the notations\begin{eqnarray}
y_{+} & = & \omega z+\omega^{-1}z^{-1}\label{eq:ypmeq}\\
y_{-} & = & \omega^{-1}z+\omega z^{-1}\qquad,\qquad\omega=e^{i\pi\frac{B}{2}}\nonumber \end{eqnarray}
with the auxiliary variable $z$ defined as a solution of $y=z+z^{-1}$;
i.e. from (\ref{eq:ydef}) one has $z=\mathrm{e}^{\theta}$.

Let us define the elementary symmetric polynomials by\begin{eqnarray*}
\prod_{i=1}^{n}(x+x_{i}) & = & \sum_{l=1}^{n}x^{n-l}\sigma_{l}^{(n)}(x_{1},\dots,x_{n})\\
\sigma_{l}^{(n)}\equiv0 &  & \mbox{if }\: l<0\mbox{ or }l>n\end{eqnarray*}
The upper index will be omitted in the sequel, as the number of variables
will always be clear from the context. Let us also denote\[
[n]=\frac{\omega^{n}-\omega^{-n}}{\omega-\omega^{-1}}=\frac{\sin\frac{n\pi B}{2}}{\sin\frac{\pi B}{2}}\]
 Let us also introduce the polynomials $P_{n}^{(k)}$:\begin{eqnarray*}
P_{1}^{(k)} & = & [k]\\
P_{n}^{(k)} & = & [k]\det M^{(n)}(k)\qquad n>1\\
 &  & M_{ij}^{(n)}(k)=[i-j+k]\sigma_{2i-j}(x_{1},x_{2}\dots,x_{n})\quad i,j=1,\dots,n-1\end{eqnarray*}
which are ingredients of the bulk form factor solution \cite{koubek_mussardo}. 

In terms of these definitions, the form factor solution can be written
as \begin{equation}
Q_{n}^{(k)}=\epsilon_{1}\epsilon_{2}B_{n-1}Q_{n-1}^{(k)}+\sigma_{n}P_{n}^{(k)}+\sigma_{n}A_{n}^{(k)}\label{eq:ffansatz}\end{equation}
where $A_{n}^{(k)}$ is a linear combination of products of $\sigma_{l}$
with total degree strictly less than $n(n-1)/2$, and the first term
is understood with the replacement\[
\sigma_{l}^{(n-1)}\rightarrow\sigma_{l}^{(n)}\]
and the notation\[
\epsilon_{1}=2\cos\gamma\qquad,\qquad\epsilon_{2}=2\cos\gamma'\]
was introduced. The $A$-polynomials are given by \begin{eqnarray}
A_{2}^{(k)} & = & 0\nonumber \\
A_{3}^{(k)} & = & [k](\epsilon_{1}^{2}+\epsilon_{2}^{2}+[k]\epsilon_{1}\epsilon_{2})\sigma_{1}\nonumber \\
A_{4}^{(k)} & = & 4\sin^{2}\frac{\pi B}{2}\,[k]^{2}\sigma_{1}\sigma_{3}+[k]^{2}(\epsilon_{1}^{2}+\epsilon_{2}^{2}+[k]\epsilon_{1}\epsilon_{2})\sigma_{1}^{2}\left(\sigma_{2}+4\sin^{2}\frac{\pi B}{2}\right)\label{eq:Apolynomials}\end{eqnarray}
up to $4$-particle level, and it can easily be extended to higher
levels using any symbolic algebra software.

These form factors correspond to the field\[
\frac{1}{\left\langle \mathrm{e}^{k\frac{b}{2}\Phi(0,t)}\right\rangle }\mathrm{e}^{k\frac{b}{2}\Phi(0,t)}\]
which is normalized to have unite vacuum expectation value.

\section{Breather boundary form factors in the sine-Gordon model}

\subsection{The boundary form factors of multi-$B_{1}$ states}

The theory is defined by the following action\begin{equation}
\mathcal{A}_{\mathrm{bsG}}=\int d^{2}x\left[\frac{1}{4\pi}(\partial_{t}\phi)^{2}-\frac{1}{4\pi}(\partial_{x}\phi)^{2}-2\mu\cos2\beta\phi\right]-2\mu_{B}\int dt\cos\beta(\phi(t,0)-\phi_{0})\label{eq:actionbsg}\end{equation}
The bulk spectrum of this theory consists of a soliton-antisoliton
doublet of mass $M$, with a number of breathers (depending on the
coupling $\beta$) and their $S$-matrices are known \cite{ZZ}. For
the full boundary spectrum and associated reflection factors the interested
reader is referred to \cite{BSG} (and references therein). For the
purposes of the present work only the consideration of the first breather
$B_{1}$ is needed.

In this model, the exact expectation values of boundary exponential
operators are known \cite{boundaryvev,finitesizebsg}:\begin{equation}
\left\langle 0\left|\mathrm{e}^{ia\phi(t,0)}\right|0\right\rangle =\left(\frac{\pi\mu\Gamma(1-\beta^{2})}{\Gamma(\beta^{2})}\right)^{\frac{a^{2}}{2(1-\beta^{2})}}g_{0}(a,\beta)g_{S}(a,\beta,z,\bar{z})g_{A}(a,\beta,z,\bar{z})\label{eq:exactbexpvev}\end{equation}
where\begin{eqnarray*}
g_{0}(a,\beta) & = & \exp\left\{ \int_{0}^{\infty}\frac{dt}{t}\left[\frac{2\sinh^{2}(a\beta t)\left(\mathrm{e}^{(1-\beta^{2})t/2}\cosh(t/2)\cosh(\beta^{2}t/2)-1\right)}{\sinh(\beta^{2}t)\sinh(t)\sinh((1-\beta^{2})t)}-a^{2}\mathrm{e}^{-t}\right]\right\} \\
g_{S}(a,\beta,z,\bar{z}) & = & \exp\left\{ \int_{0}^{\infty}\frac{dt}{t}\frac{\sinh^{2}(a\beta t)\left(2-\cos(2zt)-\cos(2\bar{z}t)\right)}{2\sinh(\beta^{2}t)\sinh(t)\sinh((1-\beta^{2})t)}\right\} \\
g_{A}(a,\beta,z,\bar{z}) & = & \exp\left\{ \int_{0}^{\infty}\frac{dt}{t}\frac{\sinh(2a\beta t)\left(\cos(2zt)-\cos(2\bar{z}t)\right)}{\sinh(\beta^{2}t)\sinh(t)\cosh((1-\beta^{2})t)}\right\} \end{eqnarray*}
and\begin{equation}
\cosh^{2}\pi z=\mathrm{e}^{-2i\beta\phi_{0}}\frac{\mu_{B}^{2}\sin\pi\beta^{2}}{\mu}\qquad,\qquad\cosh^{2}\pi\bar{z}=\mathrm{e}^{2i\beta\phi_{0}}\frac{\mu_{B}^{2}\sin\pi\beta^{2}}{\mu}\label{eq:zuvir}\end{equation}
provided the operators are normalized as\begin{eqnarray*}
\mathrm{e}^{2ia\phi(x)}\mathrm{e}^{-2ia\phi(x)} & \sim & \frac{1}{\left|x-y\right|^{4a^{2}}}+\dots\\
\mathrm{e}^{ia\phi(t,0)}\mathrm{e}^{-ia\phi(t',0)} & \sim & \frac{1}{\left|t-t'\right|^{2a^{2}}}+\dots\end{eqnarray*}
which also defines the normalization of the couplings $\mu$ and $\mu_{B}$.
The coupling $\mu$ can be related to the mass $M$ of the soliton
as follows \cite{mass_scale}:\[
\mu=\frac{\Gamma(\beta^{2})}{\pi\Gamma(1-\beta^{2})}\left[M\frac{\sqrt{\pi}\Gamma\left(\frac{1}{2(1-\beta^{2})}\right)}{2\Gamma\left(\frac{\beta^{2}}{2(1-\beta^{2})}\right)}\right]^{2-2\beta^{2}}\]
The analytic continuation between the sinh-Gordon and sine-Gordon
theory in the bulk is defined by \[
\beta=\frac{ib}{\sqrt{8\pi}}\]
Under this substitution the sinh-Gordon particle's $S$-matrix (\ref{eq:sinh_Smatrix})
becomes the $S$-matrix of the first breather in sine-Gordon theory:\begin{eqnarray*}
S_{B_{1}B_{1}}(\theta) & = & \left[\xi\right]_{\theta}=\frac{\sinh\theta+i\sin\pi\xi}{\sinh\theta-i\sin\pi\xi}\\
\xi & = & -\frac{B}{2}=\frac{\beta^{2}}{1-\beta^{2}}\end{eqnarray*}
The mass of the first breather is related to the soliton mass $M$
as \[
m_{1}=2M\sin\frac{\pi\xi}{2}\]
The identification of the two models can be completed by relating
the boundary parameters. Introducing the Ghoshal-Zamolodchikov parameters
\cite{GZ} \[
z=\frac{\beta^{2}}{\pi}(\vartheta-i\eta)\quad,\quad\bar{z}=\frac{\beta^{2}}{\pi}(\vartheta+i\eta)\]
the reflection factor of the first breather can be written in the
form\begin{equation}
R_{B_{1}}^{(\alpha)}(\vartheta)=\left(\frac{1}{2}\right)_{\theta}\left(\frac{1}{2}-\frac{\xi}{2}\right)_{\theta}\left(1+\frac{\xi}{2}\right)_{\theta}\left[\frac{\xi\eta}{\pi}-\frac{1}{2}\right]_{\theta}\left[\frac{i\xi\vartheta}{\pi}-\frac{1}{2}\right]_{\theta}\label{eq:sg_b1refl}\end{equation}
where $\alpha$ denotes the boundary condition parametrized by $\eta$
and $\vartheta$. Under this identification the relation (\ref{eq:zuvir})
coincides with the relation between the GZ parameters and the Lagrangian
boundary parameters derived by Alyosha Zamolodchikov \cite{sinG_uvir,finitesizebsg}.

Comparing this to (\ref{eq:shgrefl}) one obtains the identification\[
E=\frac{2\xi\eta}{\pi}\quad,\quad F=\frac{2i\xi\vartheta}{\pi}\]
As a result, the form factors of the boundary operator\[
\mathcal{O}_{k}=\mathrm{e}^{ik\beta\phi(0,t)}\]
can be written in terms of the sinh-Gordon form factor solution (presented
in subsection \ref{sub:Boundary-form-factors}) as follows\begin{eqnarray}
F_{\underbrace{1\dots1}_{n}}^{(k)}\left(\theta_{1},\dots,\theta_{n}\right) & = & \mathcal{G}_{k}H_{n}\frac{Q_{n}^{(k)}(y_{1},y_{2}\dots,y_{n})}{\prod_{i}y_{i}\,\prod\limits _{i<j}(y_{i}+y_{j})}\prod_{i=1}^{n}r(\theta_{i})\prod_{i<j}f(\theta_{i}-\theta_{j})f(\theta_{i}+\theta_{j})\nonumber \\
\mbox{with} &  & B\rightarrow-2\xi\quad,\quad\gamma=\xi\eta-\frac{\pi}{2}\quad,\quad\gamma'=i\xi\vartheta-\frac{\pi}{2}\label{eq:sing_b1_ff}\end{eqnarray}
and \[
\mathcal{G}_{k}=\left\langle 0\left|\mathrm{e}^{ik\beta\phi(t,0)}\right|0\right\rangle \]
is the exact vacuum expectation value (\ref{eq:exactbexpvev}).

\subsection{Finite volume form factors}

We now briefly recall the formalism developed in \cite{bfftcsa} for
the description of matrix elements of boundary operators in finite
volume, specialized for the levels that consist of first breathers.

\subsubsection{Finite volume energy levels}

Introduce the bulk phase-shifts\[
S_{B_{1}B_{1}}\left(\theta\right)=\mbox{e}^{i\delta\left(\theta\right)}\]
and their boundary counterparts \begin{equation}
R_{B_{1}}^{(\alpha)}\left(\theta\right)=\mbox{e}^{i\delta^{(\alpha)}\left(\theta\right)}\label{eq:bphaseshifts}\end{equation}
where $\alpha$ denotes the boundary condition. Putting the theory
on a strip of width $L$ with boundary conditions $\alpha$ and $\beta$,
the finite volume levels can be obtained by solving the Bethe-Yang
equations \cite{bbye}:\begin{eqnarray}
Q_{j}\left(\theta_{1},\dots,\theta_{n}\right) & = & 2\pi I_{j}\label{eq:bbye}\end{eqnarray}
where the phases describing the wave function monodromy are given
by\[
Q_{j}\left(\theta_{1},\dots,\theta_{n}\right)=2m_{1}L\sinh\theta_{j}+\sum_{k\neq j}\left\{ \delta\left(\theta_{j}-\theta_{k}\right)+\delta\left(\theta_{j}+\theta_{k}\right)\right\} +\delta^{(\alpha)}\left(\theta_{j}\right)+\delta^{(\beta)}\left(\theta_{j}\right)\]
Here all rapidities $\theta_{j}$ (and accordingly all quantum numbers
$I_{j}$) are taken to be positive. The quantum numbers can be taken
ordered as $I_{1}<\dots<I_{n}$; they must all be different due to
the exclusion principle ($S(0)=-1$). The corresponding multi-particle
state is denoted by \[
\vert\{I_{1},\dots,I_{n}\}\rangle_{L}\]
and its energy (relative to the ground state) is given by\[
E_{I_{1}\dots I_{n}}(L)=\sum_{j=1}^{n}m_{1}\cosh\tilde{\theta}_{j}+O(\mathrm{e}^{-\mu L})\]
where $\left\{ \tilde{\theta}_{j}\right\} _{j=1,\dots,n}$ is the
solution of eqns. (\ref{eq:bbye}) at the given volume $L$. The Bethe-Yang
equations gives the energy of the multi-particle states to all order
in $1/L$, neglecting only finite size effects decaying exponentially
with $L$ (where $\mu$ is some finite mass scale, dependent on the
details of the spectrum).

\subsubsection{Non-diagonal matrix elements}

For non-diagonal matrix elements, it was shown in \cite{bfftcsa}
that\begin{eqnarray}
 &  & \left|\langle\{I_{1}',\dots,I_{m}'\}\vert\mathcal{O}(0)\vert\{I_{1},\dots,I_{n}\}\rangle_{L}\right|=\nonumber \\
 &  & \qquad\frac{\left|F_{\underbrace{1\dots1}_{n+m}}^{\mathcal{O}}(\tilde{\theta}_{m}'+i\pi,\dots,\tilde{\theta}_{1}'+i\pi,\tilde{\theta}_{1},\dots,\tilde{\theta}_{n})\right|}{\sqrt{\rho(\tilde{\theta}_{1},\dots,\tilde{\theta}_{n})\rho(\tilde{\theta}_{1}',\dots,\tilde{\theta}_{m}')}}+O(\mathrm{e}^{-\mu L})\label{eq:genffrelation}\end{eqnarray}
where \begin{equation}
\rho(\tilde{\theta}_{1},\dots,\tilde{\theta}_{n})=\det\left\{ \frac{\partial Q_{k}(\theta_{1},\dots,\theta_{n})}{\partial\theta_{l}}\right\} _{k,l=1,\dots,n}\label{eq:dosfinvol}\end{equation}
is the finite volume density of states, which is the Jacobi determinant
of the mapping between the space of quantum numbers and the space
of rapidities given by the Bethe-Yang equations (\ref{eq:bbye}).
In general, the phase conventions used for the exact form factors
and the finite volume matrix elements differ, so only the absolute
values can be compared. Evaluating the above expansion requires analytic
continuation of the form factors to complex values of $\theta$ which
can be accomplished using the formulae given in Appendix \ref{sec:Analytic-continuation-of}.

\subsubsection{Diagonal matrix elements}

In the diagonal case, there are disconnected contributions which can
be taken care of by regularizing the appropriate form factor as \begin{eqnarray}
F_{\underbrace{1\dots1}_{2n}}(\theta_{n}+i\pi+\epsilon_{n},...,\theta_{1}+i\pi+\epsilon_{1},\theta_{1},...,\theta_{n})=\label{mostgeneral}\\
\prod_{i=1}^{n}\frac{1}{\epsilon_{i}}\cdot\sum_{i_{1}=1}^{n}...\sum_{i_{n}=1}^{n}\mathcal{A}_{i_{1}...i_{n}}(\theta_{1},\dots,\theta_{n})\epsilon_{i_{1}}\epsilon_{i_{2}}...\epsilon_{i_{n}}+\dots\nonumber \end{eqnarray}
where $\mathcal{A}_{i_{1}...i_{n}}$ is a completely symmetric tensor
of rank $n$ in the indices $i_{1},\dots,i_{n}$, and the ellipsis
denote terms that vanish when taking $\epsilon_{i}\rightarrow0$ simultaneously.
The connected matrix element can be defined as the $\epsilon_{i}$
independent part of eqn. (\ref{mostgeneral}), i.e. the part which
does not diverge whenever any of the $\epsilon_{i}$ is taken to zero:
\begin{equation}
F^{c}(\theta_{1},\theta_{2},...,\theta_{n})=n!\,\mathcal{A}_{1\dots n}(\theta_{1},\dots,\theta_{n})\label{eq:connected}\end{equation}
where the appearance of the factor $n!$ is simply due to the permutations
of the $\epsilon_{i}$. The formula for the diagonal matrix element
reads 

\begin{eqnarray}
 &  & \langle\{I_{1}\dots I_{n}\}|\mathcal{O}(0)|\{I_{1}\dots I_{n}\}\rangle_{L}=\label{eq:diaggenrulesaleur}\\
 &  & \frac{1}{\rho(\tilde{\theta}_{1},\dots,\tilde{\theta}_{n})}\sum_{A\subset\{1,2,\dots n\}}F^{c}(\{\tilde{\theta}_{k}\}_{k\in A})\tilde{\rho}_{a_{1}\dots a_{n}}(\tilde{\theta}_{1},\dots,\tilde{\theta}_{n}|A)+O(\mathrm{e}^{-\mu L})\nonumber \end{eqnarray}
The summation runs over all subsets $A$ of $\{1,2,\dots n\}$. For
any such subset the appropriate sub-determinant \[
\tilde{\rho}(\tilde{\theta}_{1},\dots,\tilde{\theta}_{n}|A)=\det\mathcal{J}_{A}(\tilde{\theta}_{1},\dots,\tilde{\theta}_{n})\]
of the $n\times n$ Bethe-Yang Jacobi matrix \begin{equation}
\mathcal{J}(\tilde{\theta}_{1},\dots,\tilde{\theta}_{n})_{kl}=\frac{\partial Q_{k}(\theta_{1},\dots,\theta_{n})}{\partial\theta_{l}}\label{eq:jacsubmat}\end{equation}
can be obtained by deleting the rows and columns corresponding to
the subset of indices $A$. The determinant of the empty sub-matrix
(i.e. when $A=\{1,2,\dots n\}$) is defined to equal $1$ by convention.

Note that in contrast to (\ref{eq:genffrelation}) in (\ref{eq:diaggenrulesaleur})
it is not necessary to take the absolute value, as phase redefinitions
drop out from a diagonal matrix element.

Eqns. (\ref{mostgeneral}) and (\ref{eq:diaggenrulesaleur}) are expected
to give the finite volume matrix elements to all order in $1/L$,
neglecting only finite size effects decaying exponentially with $L$
\cite{fftcsa1,fftcsa2,bfftcsa}.

\section{Numerical verification }

The finite volume energy levels and matrix elements can be evaluated
using the boundary truncated conformal space approach (BTCSA) \cite{DPTW1}
which is an extension of the method developed by Yurov and Zamolodchikov
\cite{yurov_zamolodchikov} to boundary quantum field theories. For
the sine-Gordon model the program used here was developed in \cite{neumann,finitesizebsg},
to which the interested reader is referred for details. 

To perform a specific check of the boundary form factor solution (\ref{eq:sing_b1_ff})
we choose the operator (corresponding to $k=1$)\[
\mathcal{O}_{1}=\mathrm{e}^{i\beta\phi(0,t)}\]
and evaluate its matrix elements between the BTCSA eigenvectors numerically.
As in \cite{bfftcsa}, all energies and matrix elements are measured
in units of (appropriate powers of) the characteristic mass scale
(here given by the soliton mass $M$), and we also use the dimensionless
volume variable $l=ML$. States can be identified by matching them
with the energy levels predicted by the Bethe-Yang equations, and
the appropriate matrix elements can then be compared to the predictions
of (\ref{eq:genffrelation}) and (\ref{eq:diaggenrulesaleur}), obtained
by substituting the exact form factor solution (\ref{eq:GenAnsatz}).
The typical BTCSA cutoff value was $16$, resulting in a truncated
Hilbert space with several thousand states. Only a small, but representative
sample of our numerical results are presented; we took care to verify
our results for numerous different values of the model parameters
$\xi,\eta,\vartheta$.

\subsection{Ground state energy and vacuum expectation value}

Before embarking on the evaluation of the form factors, the accuracy
of BTCSA data can be tested by extracting the bulk and boundary vacuum
energy constants. For each bulk coupling $\xi$ we considered a number
of different boundary conditions. In all cases, on one side of the
strip $(x=L$) the boundary condition $\beta$ is a pure Neumann one\[
\eta_{0}=\eta_{N}=\frac{\pi(1+\xi)}{2\xi}\qquad\vartheta_{0}=0\]
and varied the boundary condition $\alpha$ on the other side $x=0$;
from now on we only give the parameters $\eta$, $\vartheta$ of this
other boundary condition $\alpha$ which is also where the boundary
operator $\mathcal{O}_{1}$ is located for the matrix element calculations.
As discussed in \cite{neumann}, the energy of the ground state is
predicted to be\[
E_{0}(L)=\mathcal{B}M^{2}L+\left(\mathcal{B}_{b}(\eta_{N},0)+\mathcal{B}_{b}(\eta,\vartheta)\right)M+O\left(\mathrm{e}^{-\mu L}\right)\]
where\begin{eqnarray*}
\mathcal{B} & = & -\frac{1}{4}\tan\frac{\pi\xi}{2}\\
\mathcal{B}_{b}(\eta,\vartheta) & = & -\frac{1}{2\cos\frac{\pi\xi}{2}}\left(\cos\left(\xi\eta\right)+\cosh\left(\xi\vartheta\right)-\frac{1}{2}\cos\left(\frac{\pi\xi}{2}\right)+\frac{1}{2}\sin\left(\frac{\pi\xi}{2}\right)-\frac{1}{2}\right)\end{eqnarray*}
are the bulk and boundary energy constants in units of the soliton
mass $M$. The comparison between the BTCSA results and the above
exact predictions is illustrated in table \ref{tab:Bulk-and-boundary}.

\begin{table}
\begin{centering}
\begin{tabular}{|c|c|c|c|c|c|c|}
\hline 
$\xi$ & $\eta/\eta_{N}$ & $\vartheta$ & $\mathcal{B}$ (exact) & $\mathcal{B}$ (BTCSA) & $\mathcal{B}_{b}$ (exact) & $\mathcal{B}_{b}$ (BTCSA)\tabularnewline
\hline
\hline 
50/391 & 0.5 & 0 & -0.050904 & -0.05009 & -0.33307 & -0.33238\tabularnewline
\hline 
50/311 & 0.7 & 0 & -0.064512 & -0.06350 & -0.16625 & -0.16802\tabularnewline
\hline 
50/239 & 0.9 & 0 & -0.085246 & -0.08355 & 0.04476 & 0.03796\tabularnewline
\hline 
50/311 & 0.5 & 0.2 & -0.064512 & -0.06030 & -0.33289 & -0.33204\tabularnewline
\hline 
50/391 & 0.7 & 0.5 & -0.050904 & -0.05039 & -0.17695 & -0.17767\tabularnewline
\hline
\end{tabular}
\par\end{centering}

\caption{\label{tab:Bulk-and-boundary}Bulk and boundary energy constants from
BTCSA compared to the exact predictions, with $\mathcal{B}_{b}=\mathcal{B}_{b}(\eta_{N},0)+\mathcal{B}_{b}(\eta,\vartheta)$. }

\end{table}

\begin{figure}
\begin{centering}
\psfrag{terf}{$l=ML$}\psfrag{vev}{$\langle 0| \mathcal{O}_1 | 0 \rangle_L$}
\psfrag{TCSA1}{$50/391;0.5;0$, BTCSA}
\psfrag{elmelet1}{$50/391;0.5;0$, theory}
\psfrag{TCSA2}{$2/7;0.7;0$, BTCSA}
\psfrag{elmelet2}{$2/7;0.7;0$, theory}
\psfrag{TCSA3}{$50/391;0.7;0.5$, BTCSA}
\psfrag{elmelet3}{$50/391;0.7;0.5$, theory}\includegraphics[scale=0.7]{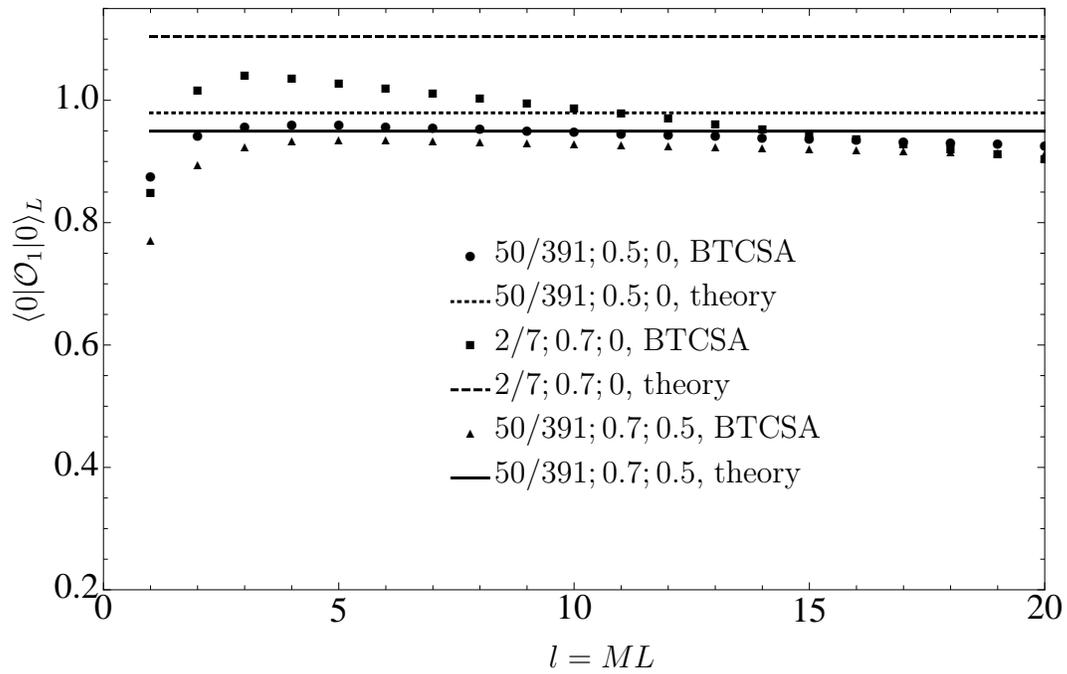}
\par\end{centering}

\caption{\label{fig:Vacuum-expectation-values:} Vacuum expectation values:
a comparison between the exact prediction and BTCSA. The model parameters
are listed as $\xi;\eta/\eta_{N};\vartheta$. All values are in units
of the soliton mass $M$. The straight lines are the infinite volume
vacuum expectation values from eqn. (\ref{eq:exactbexpvev}), the
discrete dots are the BTCSA data.}

\end{figure}
It is also possible to test the exact vacuum expectation value against
the BTCSA. The finite volume expectation value is expected to behave
as\[
\langle0|\mathrm{e}^{i\beta\phi(0,t)}|0\rangle_{L}=\mathcal{G}_{1}+O\left(\mathrm{e}^{-\mu L}\right)\]
where the exact vacuum expectation value \[
\mathcal{G}_{1}=\langle0|\mathrm{e}^{i\beta\phi(0,t)}|0\rangle\]
can be evaluated from eqn. (\ref{eq:exactbexpvev}). This is illustrated
in figure \ref{fig:Vacuum-expectation-values:}; the deviations at
large volume ($l\gtrsim5$) are due to truncation effects. Note that
agreement is better for smaller $\xi$, which agrees with the general
trend observed from all the spectral data (masses, energy levels,
bulk and boundary energy constants) that (B)TCSA converges better
if the perturbing operator is more relevant (i.e. its conformal dimension
is smaller)%
\footnote{It also happens to be the case that for a given value of $\xi$ convergence
is better for smaller values of the boundary parameter $\eta$.%
}. It is also important to notice that the derivative of the boundary
energy with respect the coupling constant $\mu_{B}$ is the expectation
value of the boundary perturbation \cite{finitesizebsg}, and so the
boundary parameter dependence of these expectation values have already
been implicitly checked by the data in table \ref{tab:Bulk-and-boundary}.

\subsection{Level identification \label{sub:Level-identification}}

\begin{figure}
\begin{centering}
\psfrag{terf}{$l$}\psfrag{ener}{$E(L)-E_0(L)$}\includegraphics[scale=0.7]{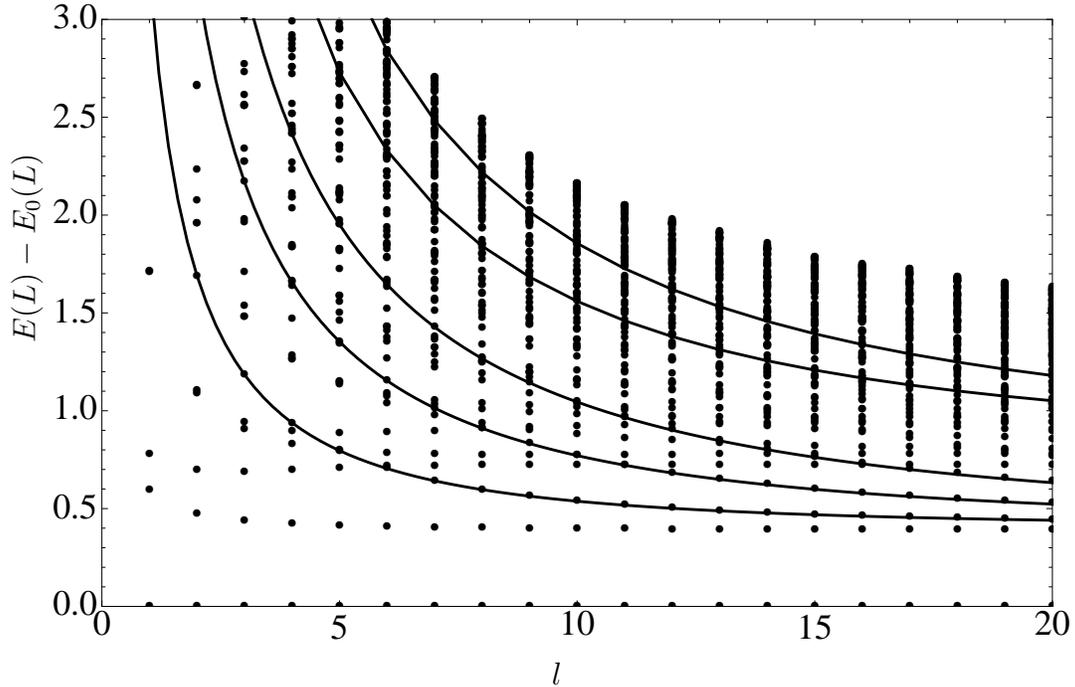}
\par\end{centering}

\caption{\label{fig:Level-identification-for} Level identification for $\xi=50/391,\eta=0.5\eta_{N},\vartheta=0$.
The dots are the BTCSA energy levels (with the vacuum subtracted),
the continuous lines are Bethe-Yang predictions for three one-particle
levels $\left|\{1\}\right\rangle $, $\left|\{2\}\right\rangle $,
$\left|\{3\}\right\rangle $ and two two-particle levels $\left|\{1,2\}\right\rangle $
and $\left|\{1,3\}\right\rangle $.}

\end{figure}

In order to evaluate matrix elements it is necessary to identify the
finite volume levels. As illustrated in figure \ref{fig:Level-identification-for},
this is performed by matching predictions from the Bethe-Yang equations
(\ref{eq:bbye}) to the BTCSA spectrum. It is apparent that this is
not an easy task, in fact, a magnitude harder than in the case of
bulk theories. In contrast to the case of periodic boundary conditions
\cite{FRT1}, there is no conserved momentum, and so the Hilbert space
cannot be split into sectors on the basis of momentum. As a result,
the continuum (in infinite volume limit) starts at the one-particle
threshold, as opposed to the bulk case. Since topological charge is
not conserved either, another way of reducing dimensionality is also
lost, resulting in an extremely dense spectrum. Additional complexity
arises from the complicated particle spectrum and the presence of
boundary bound states \cite{dirichlet,BSG} (in fact, the lowest excited
level in figure \ref{fig:Level-identification-for} is precisely such
a state and the next two are also clearly visible in the spectrum).
As a result, in contrast to the bulk theories investigated in \cite{fftcsa1,fftcsa2}
there is no way of reliably identifying states with more than two
particles. It is also apparent from the figure that the two-particle
states are already in a very dense part of the spectrum, and their
identification is made harder by the numerous level crossings characteristic
of integrable models. At certain values of the volume $L$ there can
be more than one BTCSA candidates for a given Bethe-Yang solution;
identification can be completed by selecting the candidate on the
basis of one of the form factor measurements, which still leaves other
matrix elements involving the state as cross-checks.

Level crossings also present a problem in numerical stability, since
in their vicinity the state of interest is nearly degenerate to another
one. Since the truncation effect can be considered as an additional
perturbing operator, the level crossings are eventually lifted. However,
such a near degeneracy greatly magnifies truncation effects on the
eigenvectors and therefore the matrix elements \cite{bfftcsa}.

\subsection{The one-particle form factor}

\begin{figure}
\begin{centering}
\psfrag{rap}{$\theta$}\psfrag{vOb1}{$|F_1(\theta)|$}\includegraphics[scale=0.55]{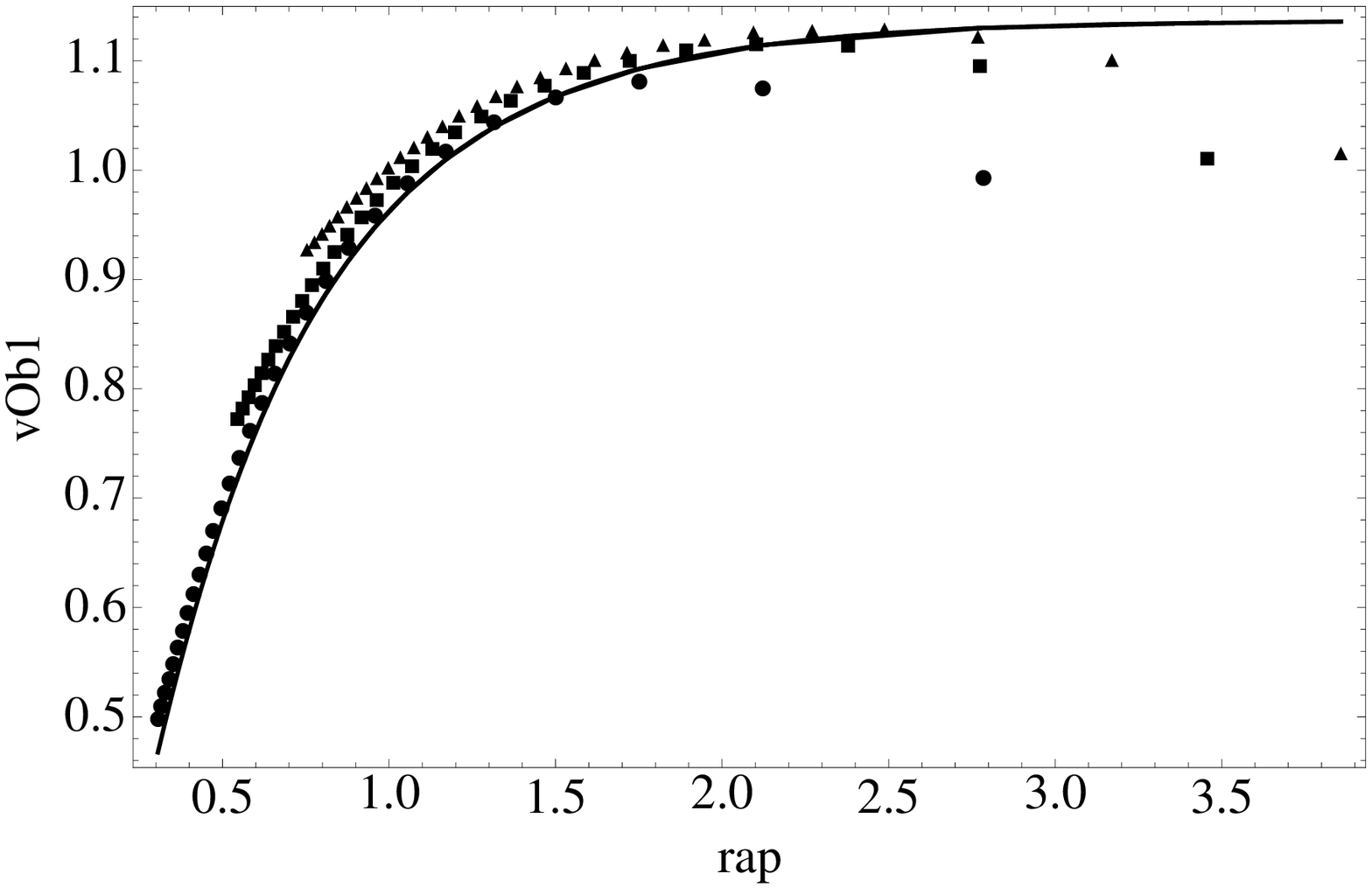}\\
\includegraphics[scale=0.55]{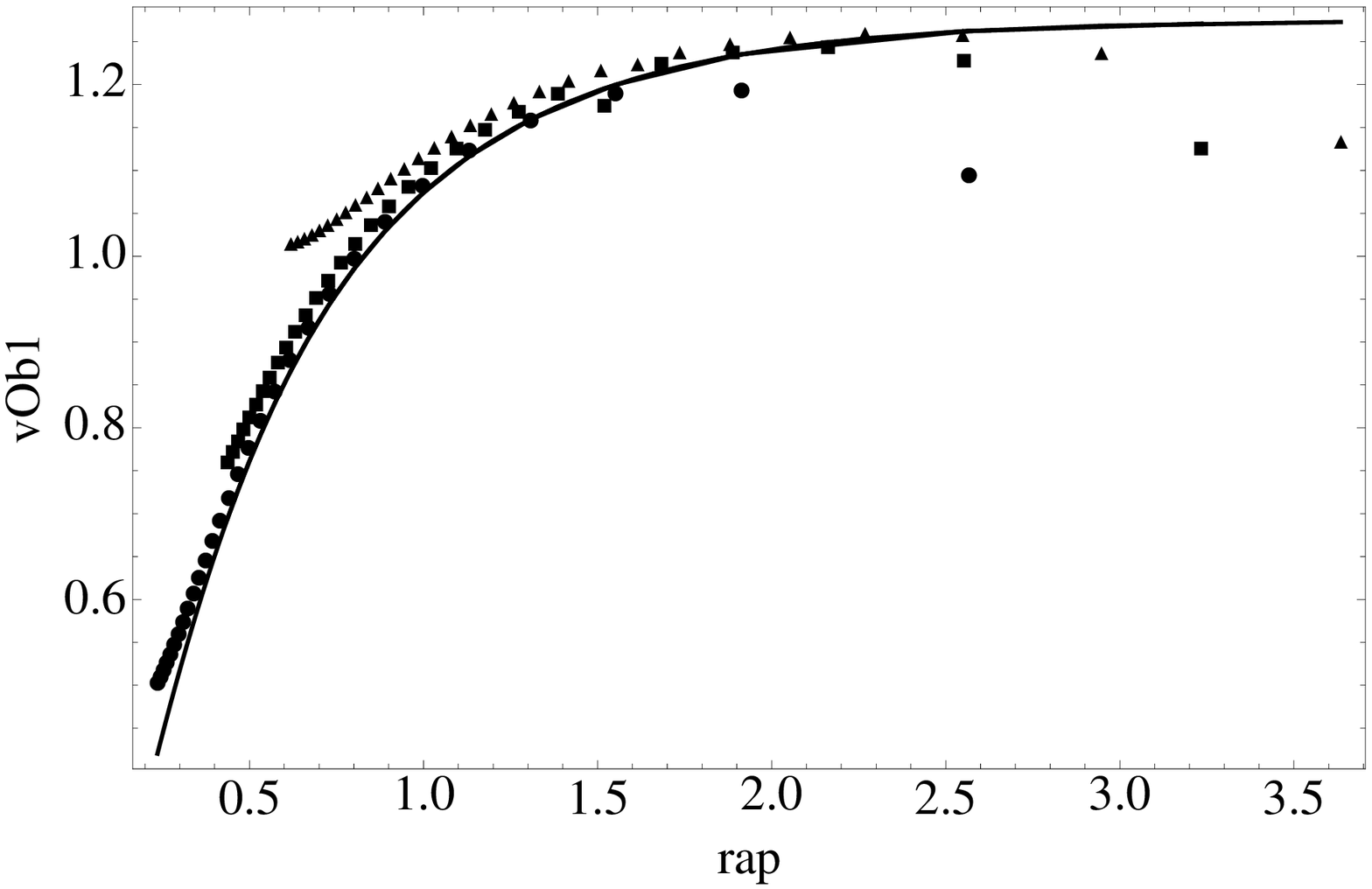}\\
\includegraphics[scale=0.55]{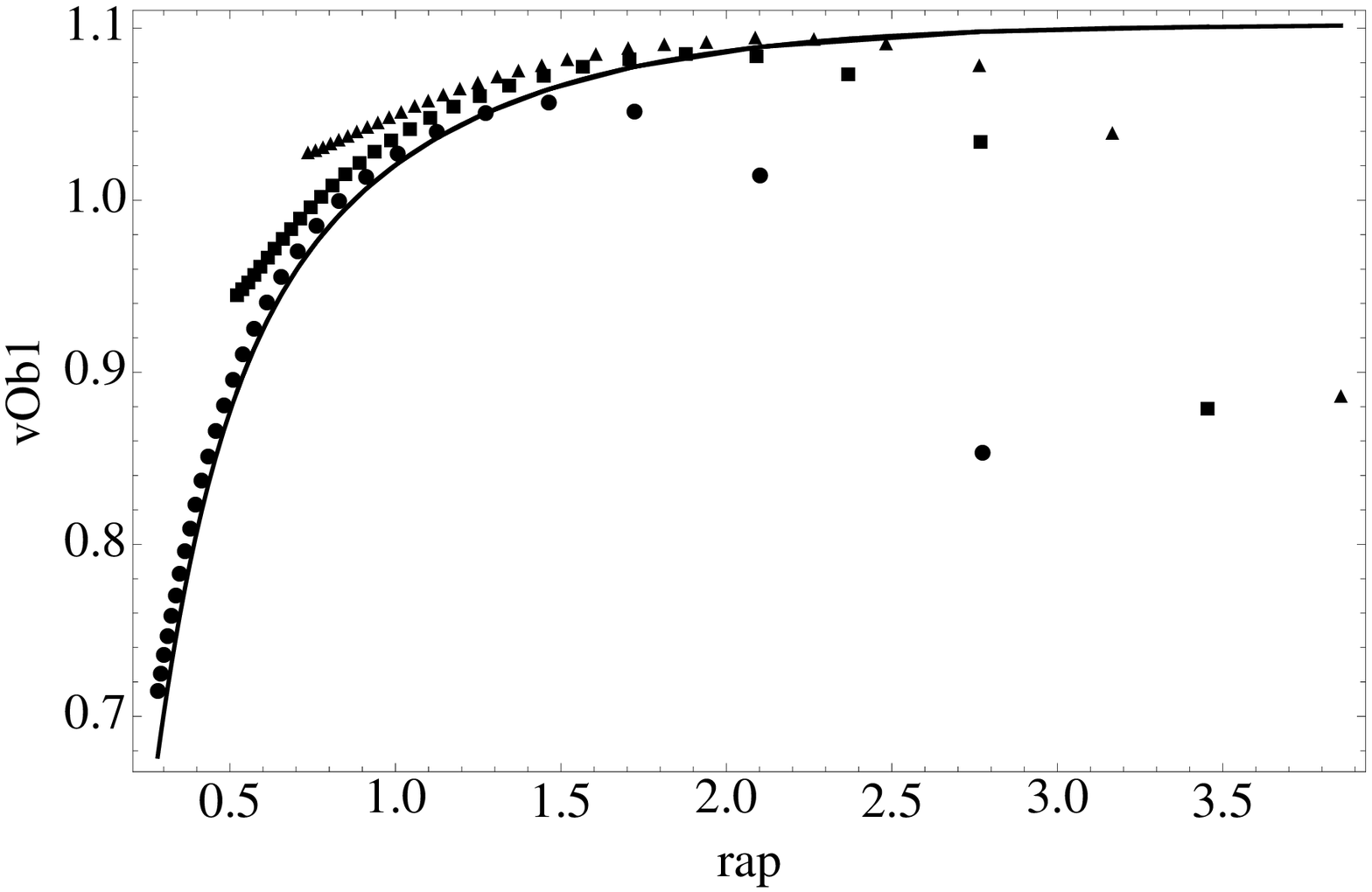}
\par\end{centering}

\centering{}\caption{\label{fig:One-particle-form-factors}One-particle form factors extracted
from BTCSA and compared to the bootstrap prediction. Continuous lines
are the bootstrap predictions, while the circles, squares and triangles
show data extracted using $I=1,2,3$ one-particle lines respectively.
The parameters $(\xi,\eta/\eta_{N},\vartheta)$ for the three figures
are: $(50/391,0.5,0)$, $(50/311,0.5,0.2)$ and $(50/391,0.7,0.5)$. }
\end{figure}

One of the most important ingredients of the boundary form factor
bootstrap is the minimal boundary form factor (\ref{eq:rmin}) which
can be obtained via measuring the one-particle matrix elements\begin{eqnarray*}
F_{1}(\theta) & = & \langle0|\mathrm{e}^{i\beta\phi(0,t)}|B_{1}(\theta)\rangle\\
 & = & \mathcal{G}_{1}\, H_{1}r(\theta)\end{eqnarray*}
(note that the polynomial $Q_{1}(y)=P_{1}(y)=[k]y=y$ for $k=1$).
Eqn. (\ref{eq:genffrelation}) can be used to express\[
\left|F_{1}(\tilde{\theta})\right|=\sqrt{\rho_{1}(\tilde{\theta})}\left|\langle0\vert\mathcal{O}(0)\vert\{I\}\rangle_{L}\right|+O(\mathrm{e}^{-\mu L})\]
where $\tilde{\theta}$ solves the Bethe-Yang equation\begin{equation}
Q_{1}(\theta)=2m_{1}L\sinh\theta+\delta^{(\alpha)}\left(\theta\right)+\delta^{(\beta)}\left(\theta\right)=2\pi I\label{eq:onept_by}\end{equation}
and \begin{equation}
\rho_{1}(\tilde{\theta})=Q_{1}'(\tilde{\theta})\label{eq:onept_bydet}\end{equation}
Therefore it is possible to compare the values extracted from several
different one-particle lines (distinguished by the value of $I$)
on the same plot. The numerics and the theoretical prediction are
in excellent agreement, as shown in figure \ref{fig:One-particle-form-factors}.
Note that the numerics deviate from the prediction for small $\theta$
(large $L$) due to truncation errors, while for large $\theta$ (small
$L$) the exponential corrections show up. The advantage of lower
$I$ states is that the scaling regime corresponds to the low-$\theta$
domain, while higher $I$ states are useful in scanning the large-$\theta$
behaviour of the form factor function.

\subsection{Two-particle form factors}

For the two-particle form factor, three independent tests can be performed:
\begin{enumerate}
\item Vacuum--two-particle matrix element:\[
\left|\langle0\vert\mathrm{e}^{i\beta\phi(0,t)}\vert\{I_{1},I_{2}\}\rangle_{L}\right|=\frac{\left|F_{11}(\tilde{\theta}_{1},\tilde{\theta}_{2})\right|}{\sqrt{\rho_{11}(\tilde{\theta}_{1},\tilde{\theta}_{2})}}+O(\mathrm{e}^{-\mu L})\]
where $\tilde{\theta}_{1},\tilde{\theta}_{2}$ are solutions of the
two-particle Bethe-Yang equations, $\rho_{11}$ is the corresponding
Bethe-Yang Jacobian, $F_{11}$ is given by the $n=2$ case of (\ref{eq:sing_b1_ff}),
and the exact $\mathcal{G}$ is inserted for proper normalization
of the operator (figure \ref{fig:Vacuum-two-particle-matrix-elements,}).%
\begin{figure}
\begin{centering}
\psfrag{terf}{$l$}
\psfrag{vacOb1b1}{$|\langle 0 | \mathcal{O}_1 | \lbrace I_1,I_2 \rbrace \rangle_{L}|$}
\psfrag{TCSA1}{$I_1=1,I_2=2$, BTCSA}
\psfrag{elmelet1}{$I_1=1,I_2=2$, theory}
\psfrag{TCSA2}{$I_1=1,I_2=3$, BTCSA}
\psfrag{elmelet2}{$I_1=1,I_2=3$, theory}\includegraphics[scale=0.7]{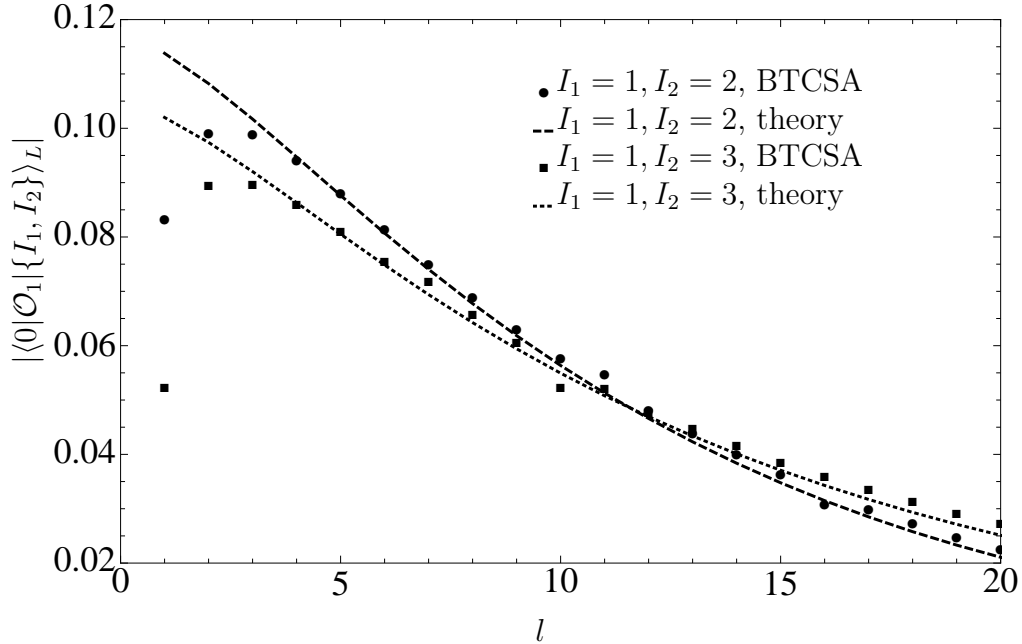}
\par\end{centering}

\caption{\label{fig:Vacuum-two-particle-matrix-elements,}Vacuum-two-particle
matrix elements for two different two-particle states, at $\xi=50/311,\eta=0.7\eta_{N},\vartheta=0$.}

\end{figure}

\item Non-diagonal one-particle--one-particle matrix element:\[
\left|\langle\{I'\}\vert\mathrm{e}^{i\beta\phi(0,t)}\vert\{I\}\rangle_{L}\right|=\frac{\left|F_{11}(i\pi+\tilde{\theta}',\tilde{\theta})\right|}{\sqrt{\rho_{1}(\tilde{\theta}')\rho_{1}(\tilde{\theta})}}+O(\mathrm{e}^{-\mu L})\]
where $\tilde{\theta}$ and $\tilde{\theta}'$ are solutions of the
one-particle Bethe-Yang equations (\ref{eq:onept_by}) with quantum
numbers $I$ and $I'$ and $\rho_{1}$ is the one-particle Bethe-Yang
Jacobian (\ref{eq:onept_bydet}) (figure \ref{fig:Non-diagonal-one-particle--one-particle-matrix}).%
\begin{figure}
\begin{centering}
\psfrag{terf}{$l$}
\psfrag{b1Ob1od}{$|\langle \lbrace I' \rbrace | \mathcal{O}_1 | \lbrace I \rbrace \rangle_L|$}
\psfrag{TCSA1}{$I'=1,I=2$, BTCSA}
\psfrag{elmelet1}{$I'=1,I=2$, theory}
\psfrag{TCSA2}{$I'=1,I=3$, BTCSA}
\psfrag{elmelet2}{$I'=1,I=3$, theory}
\psfrag{TCSA3}{$I'=2,I=3$, BTCSA}
\psfrag{elmelet3}{$I'=2,I=3$, theory}\includegraphics[scale=0.6]{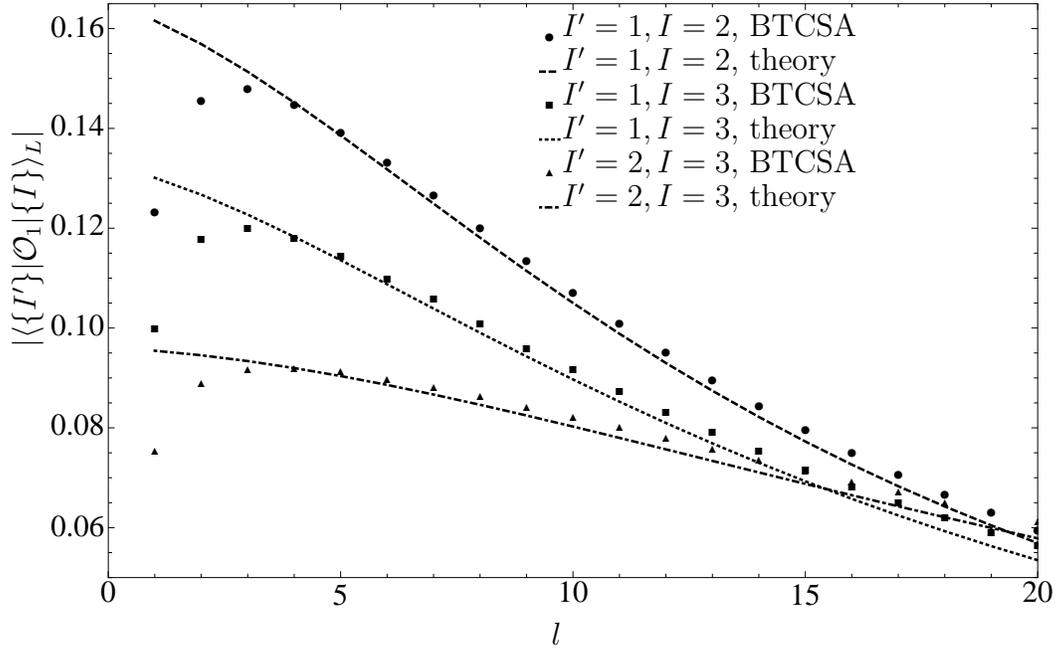}
\par\end{centering}

\caption{\label{fig:Non-diagonal-one-particle--one-particle-matrix}Non-diagonal
one-particle--one-particle matrix elements for three different choices
of the one-particle states, at $\xi=50/391,\eta=0.7\eta_{N},\vartheta=0.5$.}
\end{figure}

\item Diagonal one-particle--one-particle matrix element ($I=I'$ case):\[
\langle\{I\}\vert\mathrm{e}^{i\beta\phi(0,t)}\vert\{I\}\rangle_{L}=\frac{F_{11}(i\pi+\tilde{\theta},\tilde{\theta})}{\rho_{1}(\tilde{\theta})}+\mathcal{G}_{1}+O(\mathrm{e}^{-\mu L})\]
(figure \ref{fig:Diagonal-one-particle--one-particle-matrix}).%
\begin{figure}
\begin{centering}
\psfrag{terf}{$l$}
\psfrag{b1Ob1d}{$\langle \lbrace I \rbrace | \mathcal{O}_1 | \lbrace I \rbrace \rangle_L$}
\psfrag{TCSA1}{$I=1$, BTCSA}
\psfrag{elmelet1}{$I=1$, theory}
\psfrag{TCSA2}{$I=2$, BTCSA}
\psfrag{elmelet2}{$I=2$, theory}
\psfrag{TCSA3}{$I=3$, BTCSA}
\psfrag{elmelet3}{$I=3$, theory}\includegraphics[scale=0.65]{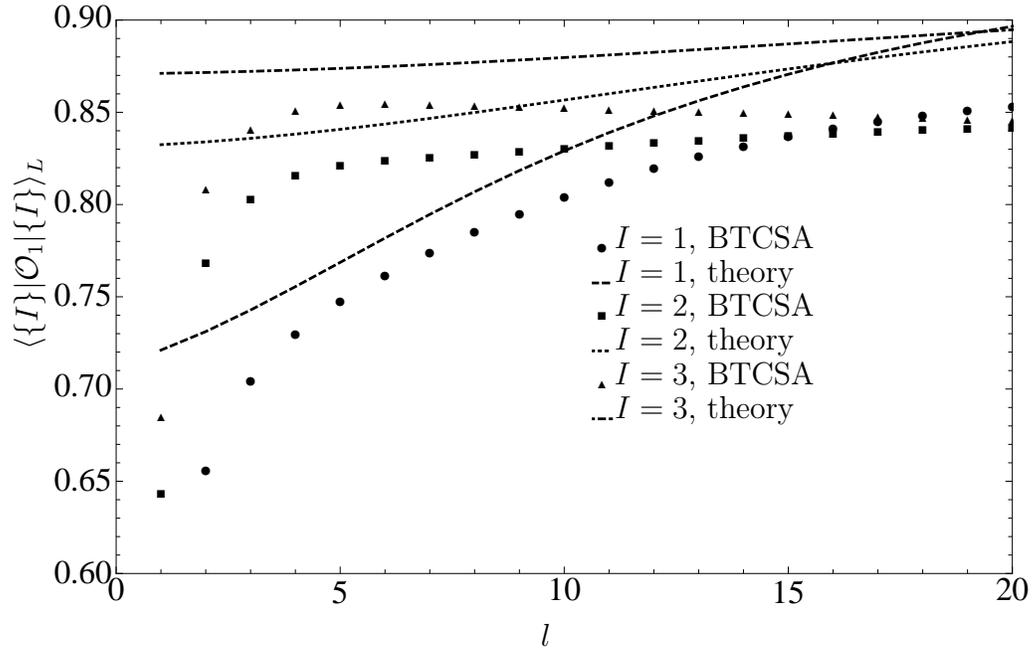}
\par\end{centering}

\caption{\label{fig:Diagonal-one-particle--one-particle-matrix}Diagonal one-particle--one-particle
matrix elements for three different choices of the one-particle state,
at $\xi=50/391,\eta=0.7\eta_{N},\vartheta=0.5$.}

\end{figure}

\end{enumerate}
The first two tests show excellent agreement%
\footnote{Even so, some points are slightly displaced. As discussed in subsection
\ref{sub:Level-identification}, this is due to the occurrence of
line crossings at the particular value of the volume, where identification
of the state becomes more difficult and BTCSA cutoff errors are also
magnified.%
}, while the third one reveals a striking discrepancy. Since this disagreement
is only seen in diagonal matrix elements, we think that the exact
form factors are correct, and the issue is with the finite size corrections
in diagonal matrix elements. A detailed discussion is given in the
conclusions.

\subsection{Higher form factors}

\begin{figure}
\begin{centering}
\psfrag{terf}{l}\psfrag{b1Ob1b1}{|$\langle \lbrace I' \rbrace | \mathcal{O} | \lbrace I_1,I_2 \rbrace \rangle_{L}|$}
\psfrag{TCSA1}{$I'=1,I_1=1,I_2=2$, BTCSA}
\psfrag{elmelet1}{$I'=1,I_1=1,I_2=2$, theory}
\psfrag{TCSA2}{$I'=2,I_1=1,I_2=2$, BTCSA}
\psfrag{elmelet2}{$I'=2,I_1=1,I_2=2$, theory}
\psfrag{TCSA3}{$I'=3,I_1=1,I_2=3$, BTCSA}
\psfrag{elmelet3}{$I'=3,I_1=1,I_2=3$, theory}\includegraphics[scale=0.7]{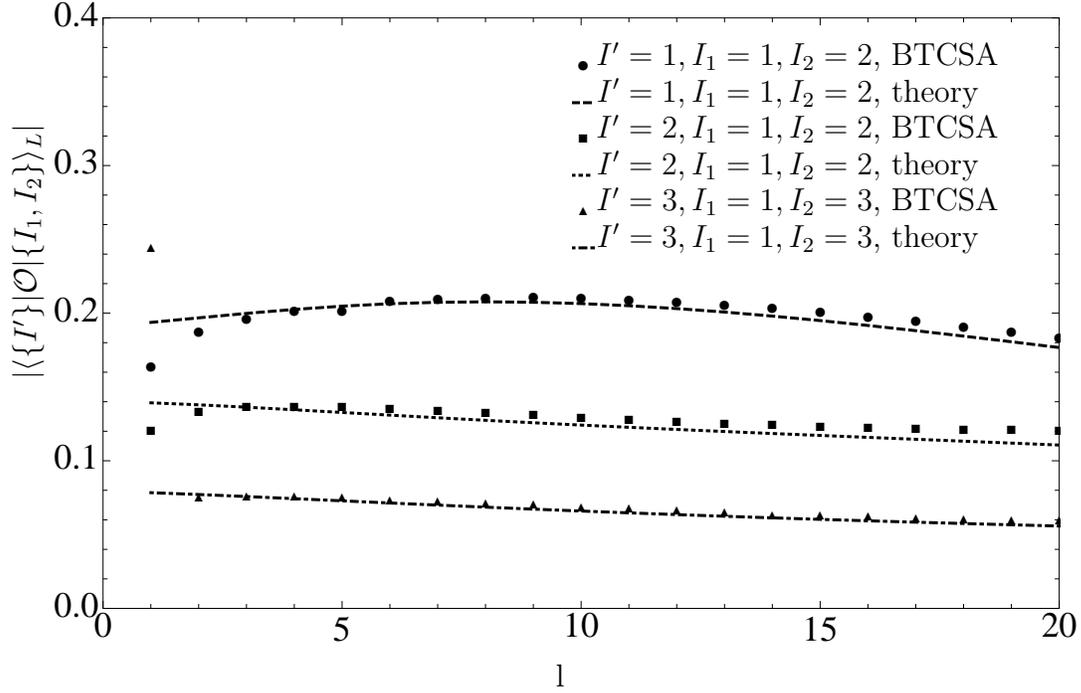}
\par\end{centering}

\caption{\label{fig:One-particle--two-particle-matrix-elements}One-particle--two-particle
matrix elements for $\xi=50/391,\eta=0.5\eta_{N},\vartheta=0$.}
\end{figure}

\begin{figure}
\begin{centering}
 \psfrag{terf}{$l$}
\psfrag{b1b1Ob1b1}{$|\langle \lbrace 1,2 \rbrace | \mathcal{O}_1 | \lbrace 1,3 \rbrace \rangle_L|$}
\psfrag{TCSA1}{BTCSA}
\psfrag{elmelet1}{theory}\includegraphics[scale=0.65]{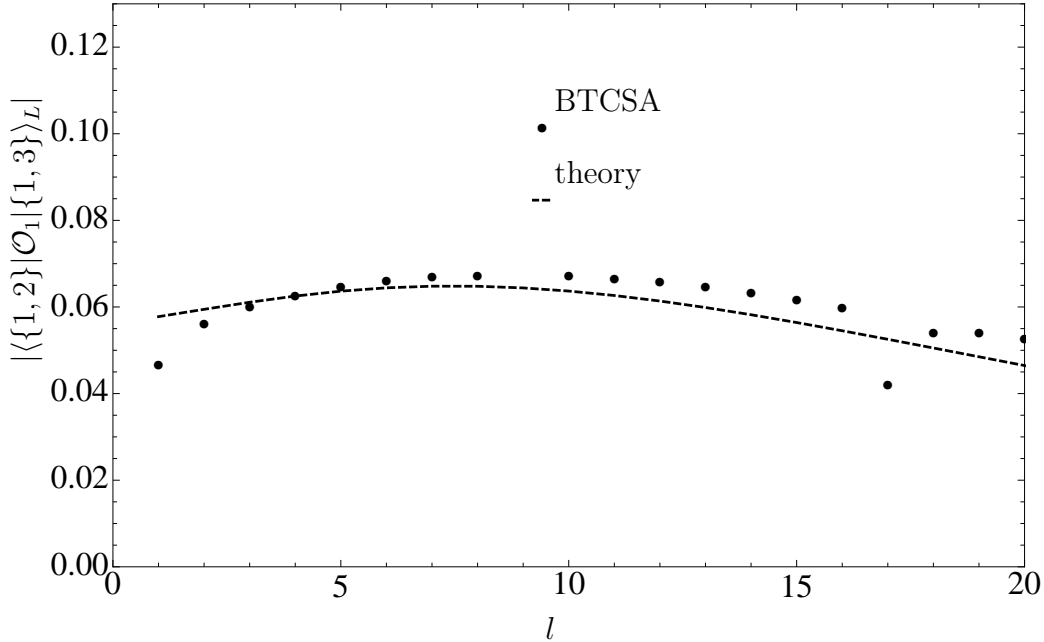}~
\par\end{centering}

\caption{\label{fig:Off-diagonal-two-particle--two-particle-matrix}Off-diagonal
two-particle--two-particle matrix elements for $\xi=50/311,\eta=0.5\eta_{N},\vartheta=0.2$.}
\end{figure}

The identified one- and two-particle states can be used to obtain
tests of the three- and four-particle form factors from (\ref{eq:sing_b1_ff}),
according to the formulae\[
\left|\langle\{I'\}\vert\mathrm{e}^{i\beta\phi(0,t)}\vert\{I_{1},I_{2}\}\rangle_{L}\right|=\frac{\left|F_{111}(i\pi+\tilde{\theta}',\tilde{\theta}_{1},\tilde{\theta}_{2})\right|}{\sqrt{\rho_{1}(\tilde{\theta}')\rho_{11}(\tilde{\theta}_{1},\tilde{\theta}_{2})}}+O(\mathrm{e}^{-\mu L})\]
and \[
\left|\langle\{I_{1}',I_{2}'\}\vert\mathrm{e}^{i\beta\phi(0,t)}\vert\{I_{1},I_{2}\}\rangle_{L}\right|=\frac{\left|F_{1111}(i\pi+\tilde{\theta}_{2}',i\pi+\tilde{\theta}_{1}',\tilde{\theta}_{1},\tilde{\theta}_{2})\right|}{\sqrt{\rho_{11}(\tilde{\theta}_{1}',\tilde{\theta}_{2}')\rho_{11}(\tilde{\theta}_{1},\tilde{\theta}_{2})}}+O(\mathrm{e}^{-\mu L})\]
(the latter is only valid for the non-diagonal case). These tests,
illustrated by figures \ref{fig:One-particle--two-particle-matrix-elements}
and \ref{fig:Off-diagonal-two-particle--two-particle-matrix}, also
show a good agreement between numerical results and theoretical expectations.
Using (\ref{eq:diaggenrulesaleur}) diagonal two-particle--two-particle
matrix elements can also be constructed, and they reveal the same
disagreement as the one-particle ones in the previous subsection.%

\section{Conclusions}

We have performed an extensive test of the form factors conjectured
in \cite{bshgff} by comparing them to results obtained using the
boundary truncated conformal space approach. For general matrix elements
an excellent agreement was found. However, for the case of diagonal
matrix elements theoretical expectations and numerical results clearly
disagree. 

The first point to make that the conjectured form factors are probably
correct, as they survived all the tests with the exception of diagonal
matrix elements. Our methods probed them up to four particles, at
which level all of the ingredients of the form factor bootstrap (minimal
form factors, recursion relations from bulk and boundary poles) are
already heavily involved. Therefore we do not expect the problem to
be related to the bootstrap.

The finite size description, on the other hand, have only been obtained
for diagonal scattering theories and the formula (\ref{eq:diaggenrulesaleur})
for the diagonal matrix elements is only an educated guess based on
the bulk case. One issue is that sine-Gordon model is a theory with
a non-diagonal scattering; however, the breather scattering is diagonal,
and therefore it is unlikely that this is the cause of the discrepancy.
The other issue is the validity of the conjectured form of disconnected
terms for the diagonal case; however, this was thoroughly tested both
in the bulk and the boundary cases \cite{fftcsa2,bfftcsa} and there
is no reason to expect any modification for sine-Gordon model.

The most likely reason for the disagreement is the presence of so-called
$\mu$-terms which can be attributed to the fact that the first breather
is a soliton-antisoliton bound states. Such effects were observed
in \cite{fftcsa1,fftcsa2} and their detailed description was given
in \cite{muterms}. It is a very interesting fact that similar effects
were found for bulk breather form factors (but not for solitonic ones!)
in sine-Gordon theory with periodic boundary conditions \cite{sgfftcsa},
which strengthens our suspicions about their origin. However, detailed
investigation of this issue requires a knowledge of solitonic form
factors of the exponential operators, which have not yet been constructed.
We intend to investigate this issue further in the near future.

\subsection*{Acknowledgments}

This work was partially supported by the Hungarian OTKA grants K75172
and K81461.

\appendix

\section{Analytic continuation of minimal form factors\label{sec:Analytic-continuation-of}}

The integral representations for the bulk (\ref{eq:fmin}) and boundary
(\ref{eq:rmin}) minimal form factors only converge in a suitable
strip around the real axis. To evaluate them at complex values with
imaginary parts as large as $\pi$, as required in (\ref{eq:genffrelation})
and (\ref{eq:diaggenrulesaleur}) they must be continued analytically
by separating pole factors that fall inside the strip of interest.
This can be accomplished using the identity

\begin{eqnarray*}
\mathcal{V}(\theta;a) & = & \exp\left[\int_{0}^{\infty}\frac{dt}{t}\left(\frac{a}{2\sinh\frac{t}{2}}-\frac{\cos\left(\frac{\theta t}{\pi}\right)\sinh at}{\sinh^{2}t}\right)\right]\\
 & = & \prod_{k=1}^{N}\left(\frac{\pi^{2}(2k-a)^{2}+\theta^{2}}{\pi^{2}(2k+a)^{2}+\theta^{2}}\right)\\
 &  & \times\exp\left[\int_{0}^{\infty}\frac{dt}{t}\left(\frac{a}{2\sinh\frac{t}{2}}-\left(N+1-N\mathrm{e}^{-2t}\right)\mathrm{e}^{-2Nt}\frac{\cos\left(\frac{\theta t}{\pi}\right)\sinh at}{\sinh^{2}t}\right)\right]\end{eqnarray*}
where the natural number $N$ is a regulatory parameter such that
the value of the functions are independent of $N$, but the width
of the convergence strip of the integral part grows with increasing
$N$.

The bulk minimal form factor can be expressed as \begin{equation}
f(\theta)=\mathcal{V}(\theta-i\pi;B/2-1)\mathcal{V}(\theta-i\pi;-B/2)\mathcal{V}(\theta-i\pi;1)\label{eq:freg}\end{equation}
and for the boundary minimal form factor function $u(\theta)$ (\ref{eq:ufunc})
one can use\begin{equation}
u(\theta)=\mathcal{U}(\theta;B/4)\mathcal{U}(\theta;1/2-B/4)\mathcal{U}(\theta;1/2)\label{eq:rreg}\end{equation}
where \begin{eqnarray*}
\mathcal{U}(\theta;a) & = & \mathcal{V}(\theta;a)\mathcal{V}(\theta-i\pi;a)\\
 & = & \exp\left[\int_{0}^{\infty}\frac{dt}{t}\left(\frac{a}{\sinh\frac{t}{2}}-\frac{2\cosh\frac{t}{2}\cos\left(\left(i\frac{\pi}{2}-\theta\right)\frac{t}{\pi}\right)\sinh at}{\sinh^{2}t}\right)\right]\end{eqnarray*}

\end{document}